\documentclass[preprint,amsmath,amssymb]{revtex4}
\usepackage{graphicx}
\usepackage{dcolumn}
\usepackage{bm}
\begin{document}
\title{Papapetrou field as the gravitoelectromagnetic field tensor in stationary spacetimes}
\author{M.~Nouri-Zonoz  \footnote{Electronic address:~nouri@ut.ac.ir, corresponding author}
and  A.~Parvizi  \footnote{Electronic
address:~a.parvizi@ut.ac.ir} } 
\address {Department of Physics, University of Tehran, North Karegar Ave., Tehran 14395-547, Iran.}
\begin{abstract}
Introducing the well known Papapetrou field as the gravitoelectromagnetic field tensor, we express the Maxwell-type part of the 3-dimensional quasi-Maxwell form of the {\it vacuum} Einstein field equations in terms of differential forms, analogous to their electromagnetic counterparts in curved spacetimes. Using the same formalism we introduce the junction conditions on non-null hypersurfaces in terms of the introduced gravitoelectromagnetic 4-vector fields and apply them to the case of the Van Stockum interior and exterior solutions.

\end{abstract}
\maketitle
%----------------------------------------------------------------------------------------
%	INTRODUCTION
%----------------------------------------------------------------------------------------
\section{Introduction}
In Einstein's general theory of relativity the old Newtonian concepts of space and time have drastically changed and fused into a new 4-dimensional, dynamical entity representing the underlying gravitational field, the so called {\it spacetime}. But when it comes to the measurements of physical phenomena, as in astrophysics and cosmology, we are led to define and measure spatial and temporal quantities and consequently a decomposition of the underlying spacetime into spatial and temporal sections would be unavoidable.
There are two well known approaches to spacetime decomposition in general relativity: the $1+3$ decomposition or {\it threading} formalism and the $3+1$ decomposition also called {\it foliation}. In the first formalism, introduced by Landau and Lifshitz \cite{Landau,Geroch,exact,lynden,Jan}, a congruence of timelike curves (world-lines) is employed for threading the spacetime under study. This is the same formalism which, through analogy with electromagnetism, enables one to introduce the so called {\it quasi-Maxwell} form of the Einstein field equations in the broader context of gravitoelectromagnetism \cite{Landau,lynden,Jan}. In the second formalism, the four-dimensional spacetime is foliated into three-dimensional hypersurfaces in the context of the so called {\it thin sandwich formulation}. This is the same formalism which led to the Hamiltonian formulation of Einstein field equations \cite{Gravitation}. In what follows we are concerned with the threading formulation of spacetime decomposition which has been quite successful in different aspects specially in finding exact solutions of Einstein field equations along with their interpretation \cite{lynden,CPM, CPM1} as well as in analyzing relativistic cosmological perturbation modes \cite{Bert}. Mathematically more involved, but essentially the same formulation of $1+3$ decomposition is presented in \cite{JHEP}, where the idea of {\it gauged motion} \cite{CPM1} was generalized to non-stationary spacetimes.\\
The analogy with electromagnetism has been employed to introduce the so called gravitoelectromagnetic (GEM) fields ${\bf E}_g$ and ${\bf B}_g$, through which the interpretation of NUT-type spacetimes as the spacetimes of gravitomagnetic monopoles was introduced \cite{lynden}. The analogy has been pushed further by introducing spacetime index of refraction \cite{Landau}, which was then used to analyze physical effects in curved backgrounds \cite {Nouri}.\\
To push this analogy even further, in the present article we show that the well known second rank antisymmetric Papapetrou field \cite{Papa, Fayos}, defined in an {\it stationary} spacetime, could be taken as the GEM field tensor. This is achieved by explicitly expressing its content in terms of the 3-dimensional gravitoelectric (GE) and gravitomagnetic (GM) fields.\\
%Obviously this second rank four-dimensional tensor field does not belong to a particular class of observers, instead it is a covariant quantity and a characteristic of the underlying stationary spacetime. 
The outline of the paper is as follows. In the next section we introduce $1+3$ decomposition and the quasi-Maxwell form of the Einstein field equations in their 3-dimensional (vector) form. In section III, introducing the Papapetrou field for stationary gravitational fields, we show explicitly how it contains the 3-dimensional GE and GM fields, hence calling it the GEM field tensor. Also we express that part of the quasi-Maxwell form of the vacuum Einstein field equations which is analogous to vacuum Maxwell equations in the language of differential forms. In the same section  we will discuss the generalized Maxwell equations in curved spacetimes and compare them with their gravitational analogues introduced in section III. In section IV the extrinsic curvature will be written in terms of the GEM tensor field and then the gravitational junction conditions are given in analogy with electromagnetism. Also, as an interesting example of the junction conditions in terms of the GE and GM fields, the case of the Van Stockum interior and exterior solutions will be discussed. In the conclusion section we  summarize and discuss our results.\\
{\bf Notations}: Following Landau and Lifshitz \cite{Landau} our convention for indices is such that Latin indices run from 0 to 3 while the Greek ones run from 1 to 3. Throughout we employ gravitational units in which $c = G =1$. Also indices ``$g$'' and ``${em}$'' stand for gravitational and electromagnetic entities respectively.
%----------------------------------------------------------------------------------------
%	FIRST SECTION: INTRODUCING GRAVITOELECTROMAGNETIC TENSOR
%----------------------------------------------------------------------------------------
\section{$1+3$ spacetime decomposition and the quasi-Maxwell form of the Einstein field equations}
Suppose that ($\mathcal{M}, g_{ab})$ is a 4-dimensional stationary spacetime/manifold with a {\it timelike killing vector field } (TKVF) $\xi\equiv\xi^a\partial_a$ representing a 1-dimensional group of transformations,
\begin{equation}
x^{a}\rightarrow x^{a}+\delta\lambda\xi^{a}    \qquad\delta\lambda\ll1,\label{sem-trans}
\end{equation}
under which the space-time line element is invariant,
\begin{equation}
\mathcal{L}_{\xi}g_{ab}=0.
\end{equation}
Obviously all vector fields $ e^c \xi ~(c\in R)$  have the same integral lines, i.e they are determined up to a constant multiplicative factor, indicating the freedom in choosing the unit of (world) time interval. The freedom in choosing the time origin at each spatial point is denoted by the following transformation \cite{Landau},
\begin{equation}
x^0 \rightarrow {x^{\prime}}^0 = (x^0 + f(x^\mu))\;\;\; , \;\;\; \mu=1,2,3 \label{trans}
\end{equation}
with $f(x^\mu)$ an arbitrary function of spatial coordinates. In the $1+3$ (threading) formulation of spacetime decomposition the line element of a stationary spacetime is written in the following form \cite{Landau,lynden,exact}
\begin{equation}
ds^2 = d\tau_{syn}^2 - dl^2 = e^{2\phi} ({A_g}_a dx^a)^2-\gamma_{\mu\nu} dx^\mu dx^\nu
\end{equation}
where $d\tau_{syn}$ is the synchronized proper time,  ${A_g}_a \equiv - g_{0a}/g_{00} = (-1, {A_g}_\alpha)$ (i.e ${A_g}_\alpha = -\frac{g_{0\alpha}}{g_{00}}$) and
\begin{equation}
dl^2 = \gamma_{\mu\nu}dx^\mu dx^\nu=(\frac{g_{0\mu} g_{0\nu}}{g_{00}} -g_{\mu\nu})dx^\mu dx^\nu \qquad , \qquad  \mu , \nu=1,2,3 
\end{equation}
is the spatial line element (also called the radar distance element) of the 3-space $\Sigma_3$ with the spatial metric $\gamma_{\mu\nu}$. Under the transformation of the time origin \eqref{trans}, as expected, the spatial line element is invariant and the spacetime metric transforms into
\begin{equation}
ds^2 = e^{2\phi} (d{x^{\prime}}^0 - {A^{\prime}_g}_{\alpha} dx^\alpha)^2-\gamma_{\mu\nu} dx^\mu dx^\nu
\end{equation}
where ${A^{\prime}_g}_{\alpha} = A{_g}_{\alpha} + \nabla_\alpha f$. In other words, the so called {\it GM potential} ${\bf A}_g$ undergoes a gauge transformation. It should be noted that the 3-space $\Sigma_3$ introduced in this formalism is the quotient space/manifold $\frac{\cal M}{G_1}$ in which $G_1$ is the one dimensional group of motions generated by the TKVF of the spacetime \cite{Geroch,exact}. This is a 3-space which does not correspond to any hypersurface embedded in the 4-dimensional spacetime as its natural habitat. One of the main advantages of the $1+3$ formulation is the fact that one could express the Einstein field equations in the so called {\it quasi-Maxwell} form in a broader context called {\it gravitoelectromagnetism}. Indeed using the above formalism, it is shown that test masses moving on the geodesics of a {\it stationary} spacetime, 
depart from the geodesics of the 3-space $\Sigma_3$ as if acted on by the following GEM Lorentz-type 3-force \cite{Landau,lynden},
\begin{equation}
{\bf f}_g = \frac{m_0}{\sqrt{1-\frac{v^2}{c^2}}}\left( {\bf E}_g + \frac{\bf v}{c} \times e^{\phi}{\bf B}_g\right)
\end{equation}
in which the 3-velocity of the particle is defined in terms of the {\it synchronized proper time} as follows \cite{Landau}
\begin{eqnarray}\label{3v}
v^\alpha = \frac{dx^\alpha}{d\tau_{syn}} = \frac{dx^\alpha}{\sqrt{g_{00}}(dx^0 - {A_\beta} dx^\beta)},
\end{eqnarray} 
and the GE and GM {\it vector fields} are defined as follows,
\begin{gather}
\textbf{B}_g = curl~({\bf A}_g) \\
\textbf{E}_g = -{\bf {\nabla}} \phi.
\end{gather}
Employing the above definitions, {\it vacuum} \footnote{Here for simplicity we restrict our attention to the vacuum case while generalization to the non-vacuum case is straightforward.} Einstein field equations could be written in the following quasi-Maxwell form \cite{lynden,Landau},
\begin{gather}
div~ \textbf{B}_g = 0 \label{hom1} \\
curl~ \textbf{E}_g=0 \label{hom2} \\
div~ \textbf{E}_g = \frac{1}{2} e^{2\phi} B_g^2+E_g^2 \label{hom3}\\
curl~(e^\phi \textbf{B}_g)=2\textbf{E}_g \times e^\phi \textbf{B}_g \label{hom4} \\
P^{\mu\nu}=-{E}_g^{\mu;\nu}+\frac{1}{2}e^{2\phi}(B_g^\mu B_g^\nu - B_g^2 \gamma^{\mu\nu})+ {E}_g^\mu E_g^\nu \label{hom5}
\end{gather}
in which $P^{\mu\nu}$ is the 3-dimensional Ricci tensor of the 3-space $\Sigma_3$ constructed from the 3-dimensional metric $\gamma^{ab}$  in the same way that the usual 4-dimensional Ricci tensor $R^{ab}$ is made out of $g^{ab}$ 
\footnote{It should be noted that the curl and divergence operators are defined in the 3-space $\Sigma_3$ with metric $\gamma_{\alpha\beta}$ in the following way;\\
%\begin{eqnarray}
$div\textbf{V}=\frac{1}{\sqrt{\gamma}}~\frac{\partial}{\partial{x^\alpha}}(\sqrt{\gamma}~V^\alpha)
~~~,~~~(curl\textbf{V})^\alpha=\frac{1}{2\sqrt{\gamma}}~\epsilon^{\alpha\beta\gamma}
(\frac{\partial{V_\gamma}}{\partial{x^\beta}}-\frac{\partial{V_\beta}}{\partial{x^\gamma}})$.\\
Also it is noted that the first two equations \eqref{hom1}-\eqref{hom2} are direct consequences of our definitions of GE and GM fields and the original ten field equations are now given by equations \eqref{hom3}-\eqref{hom5}.}.
For our later use we call the first four equations as the {\it Maxwell-type equations} and looking at the equations \eqref{hom3}-\eqref{hom4}, in analogy with electromagnetism, one could also define the following quantities;
\begin{gather}
u_g = \frac{1}{2} e^{2\phi} B_g^2 + E_g^2 \label{en1}\\
{\bf H}_g = e^{\phi} {\bf B}_g \label{en2}\\
{\bf S}_g = {\bf E}_g \times {\bf H}_g \label{en3}
\end{gather}
as the GEM energy density, the {\it GM intensity} vector and  the GEM Poynting vector respectively.\\
As an interesting observation we note that by taking the trace of the equation \eqref{hom5}, we end up with 
\begin{equation}
P = -\frac{3}{2}e^{2\phi}{B_g}^2.
\end{equation}
This equation indicates that the absence of the gravitomagnetic field (which guarantees the staticity of the underlying spacetime \cite{NKR}), lables the {\it vacuum} static solutions as those solutions with vanishing ${{\Sigma}_3}$-Ricci scalar. In other words while vacuum solutions are called Ricci flat ($R=0$), {\it static} vacuum solutions could be called ${{\Sigma}_3}$-Ricci flat solutions ($P=0$).\\ 
For future reference, two points need to be emphasized here with respect to the above quasi-Maxwell form of the Einstein field equations. The first point is the obvious fact that in terms of the GE and GM fields, equations \eqref{hom3}-\eqref{hom5} are nonlinear. As a second point, it should be noted that since the GE and GM fields are 3-vectors living in $\Sigma_3$, the above formulation is not a manifestly covariant formulation and indeed it was not meant to be so due to the idea of decomposition. In the next section introducing the  Papapetrou field as the gravitoelectromagnetic field tensor, we express the Maxwell-type part of the 3-dimensional quasi-Maxwell form of the {\it vacuum} Einstein field equations in terms of differential forms, analogous to their electromagnetic counterparts in curved spacetimes.
%%%%%%%%%%%%%%%%%%%%%%%%%%%%%%%%%%%%%%%%%%%%%%%%
\section{Papapetrou field as the Gravitoelectromagnetic field tensor of stationary spacetimes} 
Using the TKVF of the underlying stationary spacetime and in analogy with electromagnetism one can introduce the following invariant 2-form called Papapetrou field \footnote{It should be noted that the Papapetrou field could be defined for non-null Killing vectors \cite{Fayos}, but here we are interested in TKVF as their existence characterizes stationary spacetimes.}
\begin{equation}
F_g = d \xi^\star \label{2-form0}
\end{equation} 
where the 1-form $\xi^\star$ is the dual vector of the Killing vector $\xi$. 
In terms of tensor components it will take the following form,
\begin{equation}
F_g = \frac{1}{2}{F_g}_{ab} dx^{a}\wedge dx^{b},\qquad {F_g}_{ab}=\xi_{b;a}-\xi_{a;b} \label{compoF}.
\end{equation} 
It is worth noting that $F_g$ is defined on the underlying four-dimensional stationary spacetime $\cal M$, i.e the raising and lowering of its indices are done by the spacetime metric $g_{ab}$. Also it is clear from the above equation that, in analogy with the electromagnetic four-potential and field tensor, the Killing vector field $\xi^a$ and the Papapetrou field are playing the roles of a GEM 4-potential and the GEM field tensor respectively. It should be noted that in a more faithful analogy to electromagnetism it is expected that the GE and the GM potentials $\phi \equiv \ln \sqrt{g_{00}}$ and  ${\bf A}_g$, introduced in the last section, to constitute the gravitational 4-vector potential. But for the obvious reason that the GM potential is defined as a 3-vector in $\Sigma_3$, we are not able to parallel the electromagnetism here 
\footnote{Indeed one may choose ${\cal A}_a = (\phi , {\bf A}_g)$ as the GEM 4-vector potential out of which a GEM field tensor could be defined in the usual manner, namely ${{\cal F}_g}_{ab}={\cal A}_{b;a}-{\cal A}_{a;b}$.
But then the question arises as in what space these apparently 4-objects are defined? and what is the metric for raising and lowering of indices on geometrical objects in this space?. An obvious answer would be the space ${\cal R} \times \Sigma_3$ but this is clearly different from the general stationary spacetime we started from and one should present a plausible interpretation for the introduction of such a space.}.
Invariance of the GEM field tensor $F_g$ under the transformation \eqref{sem-trans} could be examined through the application of the Lie derivative on $F_g$ along $\xi$. In the coordinate system adapted to the TKVF,
\begin{equation}
\xi^a\doteq\delta_0^a=(1,0,0,0) = -g_{00} {A_g}^a,  \qquad \xi_a\doteq(g_{00},g_{0 \alpha}) = -g_{00} {A_g}_a \label{precoor}.
\end{equation}
using \eqref{compoF} and the fact that $\mathcal{L}_{\xi}\xi=0$, it is easy to see that we end up with 
\begin{equation}
\mathcal{L}_{\xi}{F_g}_{ab}=0,
\end{equation}
as a tensorial relation valid in any coordinate system proving that $F_g$ is invariant under the transformation along the TKVF. 
%----------------------------------------------------------------------------------------
%	SECOND SECTION: QUASI-MAXWELL EQUATIONS IN COVARIANT FORM
%----------------------------------------------------------------------------------------
%%%%%%%%%%%%%%%%%%%%%%%%%%%%%%%%%%%%%%%%%%%
\subsection{Maxwell-type part of quasi-Maxwell equations in the language of differential forms}
To express the Maxwell-type part of the quasi-Maxwell form of the vacuum Einstein field equations which resembles the vacuum Maxwell equations, as a first step and in analogy with the electromagnetic field tensor, we write the introduced GEM field tensor \eqref{compoF} in terms of the GE and GM 
vector fields defined in the previous section. In the coordinate system adapted to the time like Killing vector we have 
\begin{equation}
{F_g}_{ab}\doteq 2h\begin{bmatrix}
0 & {E_g}_\beta \\
-{E_g}_\alpha & 2({E_g}_{[ \alpha} {A_g}_{\beta ]}) - \frac{1}{2}f_{\alpha\beta}\\
\end{bmatrix}, \label{covcom}
\end{equation}
with its contravariant counterpart given by
\begin{equation}
{F_g}^{ab} \doteq \begin{bmatrix}
0 & -2E_g^\beta + h({\bf A}_g\times {\bf B}_g)^\beta \\
2E_g^\alpha - h({\bf A}_g\times {\bf B}_g)^\alpha & - h f^{\alpha\beta} \\
\end{bmatrix},
\end{equation}
where 
\begin{gather*}
h \equiv g_{00} \equiv e^{2\phi}\\
f_{\alpha\beta}=\sqrt{\gamma} \epsilon_{\alpha\beta\gamma} B_g^\gamma
\end{gather*}
in which $\gamma = {\rm det} \gamma_{\alpha\beta}$ and use is made of the fact that $\sqrt{-g}=\sqrt{h}\sqrt{\gamma}$ \cite{Landau}. 
Using equations  \eqref{compoF} and \eqref{covcom} one can show that , in a general coordinate system,  the GEM field $F_g$ could be written as follows
\begin{equation}
{F_g}_{ab} = - |\xi|^2 \eta_{n m  a b}\frac{\xi^n}{|\xi|} B_g^m+2(\xi_a {E_g}_b-\xi_b {E_g}_a )\label{f-deco}
\end{equation}
where $\eta_{n m  a b}= \sqrt{-g} \epsilon_{n m  a b}=\sqrt{h} \sqrt{\gamma}\epsilon_{n m  a b}$ is the 4-dimensional Levi-Civita pseudo-tensor. Conversely the GE and GM 4-vector fields are given by \cite{Nouri99}
\begin{gather}
{E_g}_a = -\frac{1}{2|\xi|} {F_g}_{ab} \frac{\xi^b}{{\vert\xi\vert}}\\
B_g^a = -\vert\xi\vert \xi^b \eta_b^{a m n} (\frac{\xi_n}{{\vert\xi\vert}^2})_{;m} = -\frac{1}{2 |\xi|^2}\eta^{a b n m} \frac{\xi_b}{\vert\xi\vert} {F_g}_{nm} \label{b}
\end{gather}
where it is obvious that ${E_g}_a\xi^a=0$ and ${B_g}_a\xi^a=0$, ensuring that these 4-vectors have no components along the TKVF, i.e in the coordinate system adapted to the Killing vector they reduce to ${E_g}_0 \doteq 0 \doteq {B_g}_0$ as expected. The above introduced GM 4-vector is proportional to the so called {\it twist} of the Killing vector field $ \omega^a =  \xi^b \eta_b^{a m n} {\xi_{n;m}} $ \cite{Geroch,Fayos}, indeed it can be shown that $B_g^a = -\frac{1}{\vert\xi\vert} \xi^b \eta_b^{a m n} {\xi_{n;m}} \equiv  -\frac{1}{\vert\xi\vert} \omega^a $ \footnote{In the formalism based on the normalized vector $\frac{\xi^a}{|\xi|}$ the GM intensity vector $H_g$ is defined so that it is twice the vorticity of the Killing observers \cite{Jan}.}. It is also noted that one should distinguish between the 3-vectors ${\bf E}_g$  and ${\bf B}_g$ living in $\Sigma_3$ and the spatial components of the 4-vectors ${E_g}_a$ and ${B_g}_a$ denoted by ${{}^{(3)}E_g}_\alpha$ and ${{}^{(3)}B_g}_\alpha$ respectively. For example, in the coordinate system adapted to the TKVF, one can obtain the following relations between the two quantities,
\begin{gather}
{E_g}_a = (E_0 , {{}^{(3)}E_g}_\alpha) \doteq (0 , {E_g}_\alpha) \\
{E_g}^a = (E^0 , {{}^{(3)}E_g}^\alpha) \doteq (-{A_g}_\alpha {E_g}^\alpha , -{E_g}^\alpha)\label{GEMM},
\end{gather}
in which it is noted that ${E_g}^\alpha = \gamma^{\alpha\beta}{E_g}_\beta$.\\
Using the above expressions the homogeneous quasi-Maxwell equations \eqref{hom1} and \eqref{hom2} could be compactly written in the following form,
\begin{equation}
\partial_{[a}{F_g}_{bc]}=0.
\end{equation}
For the two inhomogeneous equations \eqref{hom3} and \eqref{hom4}, it can be shown that they are compactly encoded in 
the following 4-dimensional form (refer to appendix A for a detailed calculation),
\begin{equation}
\nabla_a {F_g}^{ab}=0 \label{inhomo}.
\end{equation}
This result is expected since using the relation $\nabla_a \nabla_b \xi_c = R_{dabc} \xi^d$, one can obtain 
\begin{equation}
\nabla_a{F_g}^{ab}= R^{ab} \xi_a  = \frac{8 \pi G}{c^4}\left( T^{ab} - \frac{1}{2} T g^{ab}\right)\xi_a \label{inhomo2}
\end{equation}
where in the vacuum case, $R^{ab} = 0$, leads to \eqref{inhomo} \cite{Geroch,Fayos}. This also shows the straightforward  generalization to the non-vacuum case in which we replaced the Ricci tensor with the matter energy-momentum tensor and its trace using the Einstein field equations. Now the components of this equation parallel and orthogonal to the Killing vector $\xi^a$ will lead to the non-vacuum quasi-Maxwell equations \cite{lynden}.\\
It should be noted that apart from the definition given here, there are other definitions of the GEM field tensor which are all based in one way or another on the Killing field, including those based on the normalized vector field $u^a \equiv \frac{\xi^a}{|\xi|}$ given in \cite{Jan} and the one introduced by Geroch \cite{Geroch} where $F_{ab} = \nabla_{[a} \eta_{b]}$  with $ \xi^a \eta_a = 1$, so that $\eta_a =  \frac{1}{|\xi|^2}\xi_a \doteq -A_a$. In other words these three definitions are based on three different vector fields which are respectively  I- $\xi^a$, II- $\frac{1}{|\xi|}\xi^a$ and III- $\frac{1}{|\xi|^2}\xi^a$.\\
The advantage of the second  definition is that one could define gravitoelectric and gravitomagnetic 4-vectors in the spacetime manifold and introduce a
field tensor which (formally) parallels electromagnetism almost exactly \cite{Jan}. On the other hand taking the Papapetrou field as the GEM field tensor has the advantage that in some cases, such as in the Kerr-Newman geometry, it is found to be proportional to the
electromagnetic field of the same spacetime \cite{Fayos,Gravitation}. Indeed in this special case one can replace $Q$ with $2M$ to get the GEM  field tensor from its  EM counterpart.\\ 
It has already been shown that Geroch's definition of field tensor has the deficiency that it only includes the GM vector field \cite{JHEP}.\\
Using the language of differential forms and the electromagnetic field tensor, Maxwell equations could be written in a coordinate-free language \cite{nakahara}. In the same way, employing the Papapetrou field as the  GEM field tensor along with the exterior derivative operator $d$, one could express the quasi-Maxwell equations in the language of differential forms. By its definition, the GEM 2-form \eqref{2-form0} is an {\it exact} form and so by Poincare lemma it is also a closed form  and indeed the homogeneous quasi-Maxwell equations are given by 
\begin{equation}
dF_g = 0.
\end{equation}
Whereas the inhomogeneous equations are given by 
\begin{equation}
d^*F_g = 0 \label{inhomdif}
\end{equation}
with the Hodge dual of the GEM 2-form $F_g$ defined as follows,
\begin{equation}
^\star F_g= \frac{\sqrt{-g}}{4} {F_g}^{ab} \epsilon_{a b c d} dx^c \wedge dx^d
\end{equation}
in which $\epsilon$ is the Levi-Civita symbol. The above relation shows that the dual 
form $^*F_g $ is a closed 2-form.
%%%%%%%%%%%%%%%%%%%%%%%%%%%%%%%%%%%%%%%%%%
\subsection{A brief comparison with electromagnetism in curved spacetime}
Electromagnetism in curved backgrounds have been the subject of many studies specially when the curved background is a cosmological one such as a FLRW spacetime. In almost all these investigations, the electromagnetic (Faraday) field tensor is decomposed into electric and magnetic fields, with respect to the 4-velocity $u^a$ of a {\it timelike observer} ($u^au_a = 1$) as follows \cite{Lich,ellis}
\begin{equation}
{F^{em}}_{ab} =  - \eta_{n m a b} u^n  {B^m_{em}} + (u_{a} {E_b^{em}} - u_{b} {E_a^{em}}) \label{f-deco1}
\end{equation}
so that the electric and magnetic 4-vectors are given by
\begin{gather}
{E_a^{em}} = - F^{em}_{ab} u^b \\
{B_a^{em}} = \frac{1}{2} \eta_{abnm} u^b {F^{nm}_{em}} \label{EM2}.
\end{gather}
Obviously the electromagnetic fields satisfy the relations $E^{em}_a u^a=0$ and $B^{em}_a u^a=0$, ensuring their spacelike nature in the  observer's rest frame. A simple comparison of the EM equations \eqref{f-deco1}-\eqref{EM2} with their GEM counterparts \eqref{f-deco}-\eqref{b} shows that in the case of GEM, the normalized timelike vector, $\frac{\xi^n}{|\xi|}$ (also called the {\it threading vector} in the literature of the $1+3$ splitting \cite{Jan}), plays somewhat similar role to that of the comoving observer's 4-velocity in electromagnetism in a curved background. Indeed for stationary spacetimes, in the coordinate system adapted to the timelike Killing vector,  $\frac{\xi^n}{|\xi|} \doteq u^n = (1/\sqrt{g_{00}}, 0 , 0, 0)$,  i.e it is the 4-velocity of a timelike observer in the comoving frame, the so called Killing observers. \\
In brief, it should be noted that in the case of the electromagnetism in a curved background the above EM fields and the related equations (e.g Maxwell and wave equations) written in terms of them, tell us how these electromagnetic fields behave in a (stationary) curved background and do not tell us anything about the underlying spacetime. On the other hand, as pointed out previously, gravitoelectromagnetism (Einstein field equations in their quasi-Maxwell form) is a formalism which expresses the gravitational field equations in terms of GE and GM 4-vector fields introduced through $1+3$ decomposition of the underlying spacetime metric into its spatial and temporal sections. For a detailed study on the above mentioned comparison refer to 
\cite{Jan,Cost}.
%----------------------------------------------------------------------------------------
%	THIRD SECTION: JUNCTION CONDITIONS
%----------------------------------------------------------------------------------------
\section{Junction conditions}
It is well known that electromagnetic fields on boundaries separating two different media satisfy certain junction conditions. Now that we have established the analogy between Einstein field equations in stationary spacetimes and Maxwell equations, to look for further analogy, we turn our attention to the junction conditions in general relativity and try to rewrite them  in terms of the GEM fields. 
The problem of junction conditions in general relativity occurs in the study of two different spacetimes matched at their boundary which usually separates vacuum and non-vacuum solutions of Einstein field equations such as in the case of collapsing stars and thin shells of matter. The obvious question posed is:  what are the conditions to ensure that the two spacetime metrics are joined smoothly across the hypersurface $\cal S$ presenting the boundary?. There has been a lot of studies in this direction starting with the work of Darmois \cite{darmois} and followed by Lichnerowicz \cite{lich}, O`Brien and Synge \cite{obrien} and Israel \cite{israel} to name a few. In what follows we will employ the well-known 
Darmois-Israel junction conditions on thin shells  which is widely used in studies on matching different spacetimes with thin shell boundaries  in GR and cosmology. To establish our notation, first we give a brief account of their formalism. 
In the language of distributions, after dividing spacetime manifold $\mathcal{M}$ by hypersurface $\cal S$ into two regions $\mathcal{V^+}$ and $\mathcal{V^-}$ with the corresponding metrics  $g^+_{\mu\nu}$ and $g^-_{\mu\nu}$ respectively, one can express the  spacetime metric as the following distribution-valued tensor,
\begin{equation}
g_{ab}=\Theta(\ell)g^+_{ab}+\Theta(-\ell)g^-_{ab} \label{disvalu}
\end{equation}
in which $\Theta(\ell)$ is Heaviside distribution and $\ell$ is the parameter by which the congruence of geodesics piercing the hypersurface $\cal S$ orthogonally, are parameterized.\\
Denoting the spacetime coordinates by $x^a$ and those on the hypersurface by $y^i$ \footnote{Note that here the intrinsic coordinates of the hypersurface are denoted by the middle Latin indices $(i,j,..)$.} one can show that the following conditions satisfy 
\begin{equation}
[n^a]=[e^a_i]=0
\end{equation} 
in which $e^a_j=\partial{x^a} / \partial{y^i} $ and $n^a=\partial{x^a}/ \partial{\ell}$ are the tangent and normal vectors to the hypersurface respectively and the symbol [ ] stands for jump across the hypersurface. Using the above relations one can obtain the so called Darmois-Israel junction conditions for the smooth joining of the spacetime metric on the two sides of the hypersurface as follows,
\begin{gather}
[h_{ab}]= 0 \label{DI-1}\\
[K_{ij}]= 0 \label{DI-2}
\end{gather} 
in which  $h_{ab}$ and $K_{ij}$ are the so called first and second fundamental forms (or the induced metric and extrinsic curvature of the hypersurface) respectively. The first condition guarantees a singular-free connection while the second one guarantees the non-existence of matter layer on the hypersurface $\cal S$. Indeed it can be shown that the second condition gives the sufficient condition for the regularity of the Riemman tensor on the hypersurface \cite{poisson}.
%%%%%%%%%%%%%%%%%%%%%%%%%%%%%%%%%%%%%%%%%%%%%%%%%%%%%
\subsection{Junction conditions in terms of the GEM vector fields}
In what follows our main objective is to rewrite the above introduced junction conditions for non-null hypersurfaces in terms of the GEM fields of the underlying spacetime metrics and show their formal analogy with their electromagnetic counterparts. Since by second junction condition there are no matter layers on the hypersurface, it is this condition which is expected to be a gravitational analogue, in electromagnetism, of the continuity of the electric and magnetic fields on the boundary sparating two media where there are no free charges or currents.\\
We start by writing the extrinsic curvature in terms of the tangent vectors to the hypersurface and (the covariant derivative of) the normal vector to it and decompose the relation as follows  
\begin{align}
[K_{ij}]&=[n_{a;b}] e_i^a e_j^b=-[\Gamma_{ab}^c] n_c e_i^a e_j^b \label{k1} \\
&=-[\Gamma_{ab}^0] n_0 e_i^a e_j^b-[\Gamma_{ab}^\gamma] n_\gamma e_i^a e_j^b \nonumber \\
&=-[\Gamma_{00}^0] n_0 e_\alpha^0 e_\beta^0-[\Gamma_{\mu\nu}^0]n_0 e_i^\mu e_j^\nu \nonumber \\
&~~~-[\Gamma_{\gamma0}^0] n_0 e_i^\gamma e_j^0-[\Gamma_{0\gamma}^0] n_0 e_i^0 e_j^\gamma \nonumber \\
&~~~-[\Gamma_{0\eta)}^\gamma ] n_\gamma e_i^0 e_j^\eta-[\Gamma_{\eta0}^\gamma ] n_\gamma e_i^j e_j^0 \nonumber \\
&~~~-[\Gamma_{\mu\nu}^\gamma]n_\gamma e_i^\mu e_j^\nu-[\Gamma_{00}^\gamma] n_\gamma e_i^0 e_j^0 \nonumber.
\end{align}
It should be noted that there are three different kinds of indices used in the above relation. The first few Latin indices $(a,b, c)$ run from $0$ to $3$ , the Greek ones only label spatial indices and run from $1$ to $3$ and, as pointed out earlier, the few middle Latin indices $(i,j,k)$ denote the 3-dimensional intrinsic coordinates of the hypersurface $\cal S$.
The connection coefficients in the above relation could be written in terms of the metric components, which in turn, upon using the definitions of the GEM fields and after a careful manipulation of terms, could be recollected in the following compact form (detailed calculations are given in Appendix B) 
\begin{equation}
[K_{ij}]=[P_{bc}^a] n_a e_i^b e_j^c-[\lambda_{\alpha\beta}^\gamma]n_\gamma e_i^\alpha e_j^\beta \label{k2}
\end{equation}
where
\begin{equation}
P_{bc}^a= \frac{1}{2h} [(E_g^a \xi_b-{F_g}_b^a)~\xi_c  + (E_g^a \xi_c-{F_g}_c^a)~\xi_b]
\end{equation}
and $\lambda_{\alpha\beta}^\gamma$ are the components of the three-dimensional Christoffel symbol constructed from the 3-metric $\gamma_{\alpha\beta}$. Now using the above relation, the second junction condition leads to the following two conditions
\begin{gather}
[P_{bc}^a] n_a e_i^b e_j^c = 0 \label{p} \\
[\lambda_{\alpha\beta}^\gamma]n_\gamma e_i^\alpha e_j^\beta = 0 \label{2-d ex}
\end{gather}
the second of which is just the formal three dimensional analogue (in space $\Sigma_3$ with metric $\gamma_{\alpha\beta}$) of the original condition \footnote{It is called a formal three dimensional analogue since neither $\Sigma_3$ space is a hypersurface of the underlying manifold nor the spatial components of the normal to the boundary hypersurface necessarily constitute a normal vector to  $\Sigma_3$.}
 while the first condition reduces to the following two conditions on the GEM fields
\begin{gather}
[{E_g}_a] n^a = 0 \label{e-jun} \\
[{F_g}_b^a ] n_a e_i^b = 0 \label{b-jun},
\end{gather}
where we have used the fact that $[\xi_a]=0$, i.e the TKVF of the underlying stationary spacetimes has no jump across the hypersurface $\cal S$. This could be shown by noting that the Killing vector at the boundary could be written in terms of the normal and tangent vectors to the hypersurface which satisfy \eqref{DI-1} and \eqref{DI-2}. The first equation shows that the normal component of the GE field is continuous across the hypersurface. To see the content of the second equation in a more familiar form, we substitute for the field tensor from \eqref{f-deco}, upon which, one can easily see that it reproduces condition \eqref{e-jun} as well as the following two conditions,
\begin{gather}
[{E_g}_b]e^b_i = 0 \label{f-jun} \\
\eta_{n m  a b}{\xi^n} n^a [B_g^m]e^b_i = 0 \label{g-jun}.
\end{gather}
It should be noted that the two conditions \eqref{e-jun} and \eqref{f-jun} are not independent and using the orthogonality relation $n_a e_i^a = 0$, one could be derived from the other. In other words continuity of the normal components of the GE field across the hypersurface guarantees the continuity of its tangential components and vice versa.\\
In the coordinate system adapted to the timelike killing vector (i.e $\xi^a \doteq (1,0,0,0)$), equations \eqref{e-jun}, \eqref{f-jun} and \eqref{g-jun} could be rewritten in the following more familiar 3-dimensional forms
\begin{gather}
[{E_g}_a] n^a  \doteq [{{\bf E}_g}] \cdot {\bf n} = 0 \\
[{E_g}_b]e^b_i  \doteq [{{\bf E}_g}] \cdot {\bf e}_i = 0 \\
\sqrt{\gamma}\epsilon_{\beta\alpha\mu}n^\alpha[\sqrt{h}B_g^\mu]e^\beta_i \equiv {\bf n} \times [{\bf H}_g] \cdot {\bf e}_i \doteq 0  \Rightarrow  {\bf n} \times [{\bf H}_g] \doteq 0 \label{c-jun}
\end{gather}
where they show that the tangential components of the GE field  and gravitomanetic intensity as well as the normal components of the GE field are all continuous across the hypersurface separating the two regions of the spacetime. An intersting example of the above junction conditions, discussed in the following subsection, is its application to the Van Stockum solution \cite{Van}.
%%%%%%%%%%%%%%%%%%%%%%%%%55
\subsection{Junction conditions for the interior and exterior Van Stockum solutions}
To exemplify the employment of the junction conditions in terms of the GEM fields, we study the Van Stockum solution corresponding to an interior solution of a rigidly rotating dust cylinder of radius $R$, matched to three different exterior solutions \cite{Bon}. We start by writing the metric in its  1+3-decomposed form in cylindrical coordinates $x^i \equiv (t, r, z, \phi)$ as follows, 
\begin{equation}\label{VS1}
ds^2= F(dt-\frac{M}{F}d\phi)^2-H(dr^2+dz^2)-\frac{r^2}{F}d\phi,
\end{equation}
in which $H, M$ and $F$ are functions of $r$ given by the following expressions for the  interior and exterior solutions,\\
\textit{\bf Interior}, $0\leq r \leq R$
\begin{equation}
H= e^{-a^2r^2}, \quad M=ar^2, \quad F=1 
\end{equation}

\textit{\bf Exterior}, $R \leq r$
\\ Case $ I $ : $aR<\frac{1}{2}$

\begin{align}
&H= e^{-a^2R^2}(\frac{R}{r})^{2a^2R^2}, \quad M=r \sinh(\epsilon+\theta) \sinh^{-1} 2\epsilon,\\ &F=rR^{-1}\sinh(\epsilon-\theta)\sinh^{-1}\epsilon, 
\end{align}
in which $a$ is a positive constant corresponding to the angular velocity of the fluid on the rotation axis and also $\theta=(1-4a^2R^2)^{\frac{1}{2}}\log(\frac{r}{R})$  and $\tanh\epsilon=(1-4a^2R^2)^{\frac{1}{2}} $.\\
Case $ II $ : $aR=\frac{1}{2}$
\begin{align}
&H= e^{-\frac{1}{4}}(\frac{R}{r})^{\frac{1}{2}}, \quad M=\frac{1}{2}r [1+log(\frac{r}{R}], \\ &F=rR^{-1}[1-\log(\frac{r}{R})], 
\end{align}
Case $ III $ : $aR>\frac{1}{2}$

\begin{align}
&H= e^{-a^2R^2}(\frac{R}{r})^{2a^2R^2}, \quad M=r \sin(\epsilon+\theta) \sin^{-1}2\epsilon,\\ &F=rR^{-1}\sin(\epsilon-\theta)\sin^{-1}\epsilon, 
\end{align}
where $\theta=(4a^2R^2-1)^{\frac{1}{2}} \log(\frac{r}{R})$ and  $\tan\epsilon=(4a^2R^2-1)^{\frac{1}{2}} $.
Now we can employ the GEM junction conditions which in the first part require the continuity of the normal component of the GE and tangential components of gravitomegnetic fields. Employing the following normal and tangential vectors on the boundary $r=R$, in the  coordinate system of\eqref{VS1}, 
\begin{equation}
n = (0,1,0,0), \quad e^i_a=\frac{\partial x^a}{\partial y^i}, 
\end{equation}
with $y^i=(t, z, \phi)$ denoting the coordinates on the boundary hypersurface $r=R$, the GEM junction conditions are concluded as follows.\\
\textit{A : Junction conditions on GE fields}\\
Since $F =1$ in \eqref{VS1} for the interior solution, the interior GE field is zero everywhere. On the other hand for the GE field in the  exterior solutions we have,\\
Case I:
\begin{equation}
E_r=\frac{-1}{2r}[1-\frac{\tanh(\epsilon)}{\tanh(\epsilon-\theta)}] \label{EI}.
\end{equation}
Case II:
\begin{equation}
E_r=\frac{-1}{2}[\frac{1}{r}-\frac{1}{RF}] \label{EII}.
\end{equation}
Case III:
\begin{equation}
E_r=\frac{-1}{2r}[1-\frac{\tan(\epsilon)}{\tan(\epsilon-\theta)}] \label{EIII}.
\end{equation}
It is not difficult to see that  all the above expressions vanish on the boundary as required by the continuity of the normal component of the GE field.\\
\textit{A : Junction conditions on GM fields}\\
Since the GM potential $A_\phi= \frac{M}{F}$ of the spacetime \eqref{VS1} depends only on $r$, it will possess  GM fields along the $z-$ direction. The interior gavitomagnetic field on the boundary (i.e at $r=R$) is given by $ B^z_g = -2a/e^{-a^2R^2}$ whereas  for the exterior solutions we have,\\
Case I:
\begin{equation}
B^z_g=\frac{-R\sqrt{F} }{r^2H}\frac{\sinh(\epsilon)^2 \cosh^{-1}\epsilon}{\sinh(\epsilon-\theta)^2} \label{BI}.
\end{equation}
Case II:
\begin{equation}
B^z_g=\frac{-R\sqrt{F} }{r^2H}\frac{1}{(1-\log(r/R))^2} \label{BII}.
\end{equation}
Case III:
\begin{equation}
B^z_g=\frac{-R\sqrt{F} }{r^2H}\frac{\sin(\epsilon)^2 \cos^{-1}\epsilon}{\sin(\epsilon-\theta)^2} \label{BIII}.
\end{equation}
Again, it is a straightforward calculation to see that all of the above fields reduce to the same interior field on the boundary.\\
It can be shown that the second part of the GEM junction conditions on Van Stockum solutions requires the continuity of the three components $\lambda^1_{11}, \lambda^1_{22}$ and $\lambda^1_{33} $ of the Christoffel symbols of their corresponding  3-manifold $\Sigma_3 $. Their continuity across the boundary hypersurface are examined in the following, 
\begin{equation*}
^{interior} \lambda^1_{11} =-a^2r \stackrel{r=R}{=}\; ^{exterior} \lambda^1_{11} \begin{cases}
Case \;I &-a^2R^2/r\\
Case \;II &-1/4r \\
Case \;III & -a^2R^2/r,
\end{cases} 
\end{equation*}

\begin{equation*}
^{interior} \lambda^1_{22} =a^2r \stackrel{r=R}{=}\; ^{exterior} \lambda^1_{22} \begin{cases}
Case \;I &a^2R^2/r\\
Case \;II &1/4r \\
Case \;III & a^2R^2/r,
\end{cases}  
\end{equation*}

\begin{equation*}
^{interior} \lambda^1_{33} =-r/e^{-a^2R^2} \stackrel{r=R}{=}\; ^{exterior} \lambda^1_{33}\begin{cases}
Case \;I &-\frac{R \tanh(\epsilon)}{2H} \frac{\sinh(2\epsilon-\theta)}{\sinh(\epsilon-\theta)^2} \\ \\
Case \;II & \frac{r^2}{2H} \frac{\log(r/R)-2}{RF^2} \\\\
Case \;III  &-\frac{R \tan(\epsilon)}{2H} \frac{\sin(2\epsilon-\theta)}{\sin(\epsilon-\theta)^2}.
\end{cases} 
\end{equation*}
It should be noted that the case II is characterized by $aR = 1/2$. In \cite{Mena} using a slightly different notation, the authors introduce what we have called the GM intensity (denoted by $H$) and arrive at the result that it should be continuous across a  hypersurface in the $\Sigma_3$ space (which as previously noticed is not in general a submanifold of the original spacetime manifold) which was then pulled back to the original spacetime manifold by the inverse projection map. In our case we have directly applied the junction conditions on a hypersurface in the original spacetime manifold $\cal M$  which were then expressed in terms of the GEM filed vectors. The two approaches arrive at the same result, since the continuity of the gravitomagnetic intensity is also true in our case if there are no surface stress-energy tensor on the separating hypersurface, in which case, the normal component of the gravitomagnetic intensity (and also the GM field $\bf{B}_g$), would be automatically continuous across the hypersurface. Although no explicit interior solution was given, the authors have examined the junction conditions on different possible stationary interior and exterior solutions of Einstein field equations including the (exterior) cylindrical NUT space introduced in \cite{CPM}. \\
It should be mentioned that one could also employ the $3+1$-splitting (or foliation) of spacetimes to write junction conditions across timelike hypersurfaces. This is done in the context of the so called black hole membrane paradigm \cite{Thor} for the stretched horizon which replaces the true horizon \footnote{It should be noted that in the study carried out in \cite{Thor}, the GEM formalism and quasi-Maxwell equations are only treated in the weak field limit of the Kerr black hole.}. The modern version of this paradigm using an action principle is given in \cite{pari}.
%%%%%%%%%%%%%%%%%%%%%%%%%%%%%%%%%%%%%%%%%%%%%%%%%%%%%%%%%%%%%%%%%%%%%%%55
\section{Discussion and summary}
In the present study, first we introduced the well known Papapetrou field as the GEM field tensor for stationary spacetimes and  then showed its content in terms of GE and GM fields explicitly in the coordinate system adapted to the TKVF of the underlying spacetime. In this identification, unlike the previous studies \cite{JHEP}, The GEM field tensor included both GE and GM vector fields and was also different from the one introduced by Geroch in \cite{Geroch}. Using the same field, the Maxwell-type part of the vacuum Einstein field equations were written in a differential form analogous to their EM counterparts in curved backgrounds. This is done by elevating the GEM 3-vectors ${\bf E}_g$ and ${\bf B}_g$ to 4-vectors  using the TKVF  of the spacetime. The close analogy of this formalism to electromagnetism in curved backgrounds, was shown by writing the GEM field tensor in terms of the introduced GEM 4-vector fields and the TKVF of the underlying stationary spacetime. Using this analogy, the junction conditions on a hypersurface separating two regions of a spacetime with different metrics are given in terms of the GEM 4-vector fields. Obviously, at the end these junction conditions in a given coordinate system are mathematically equivalent to those obtained in the formal description of junction condition we started with, but it is expected that one could  employ them on the basis of their analogy to the electromagnetic counterparts to look for interior solutions of the  well known exterior solutions such as in the case of Kerr spacetime \cite{nzp}, in the same way they were employed to find (recover) new (old)  solutions of Einstein field equations such as the NUT-type spacetimes \cite{lynden,CPM}. Indeed it has already shown that studying principal directions of the Papapetrou field and the Ernst potential associated with the TKVF  could  serve as a tool to look for exact solutions \cite{Fayos}. This is no coincidence as the real and imaginary components of the exterior derivative of the Ernst potential are proportional to the GE and GM fields respectively. The example of the Van Stockum solution, serving as a guide, shows the possible challenges that one might face in using the approach based on the junction conditions (in terms of the GEM fields) to look for new exact solutions.\\
It should also be noted that the employment of the threading decomposition of stationary spacetimes in the above formulation of the junction conditions, restricts its application to stationary spacetimes and its generalization to non-stationary spacetimes should employ a consistent generalization of the threading decomposition to non-stationary spacetimes such as that given in  \cite{Zelmanov,Cattan} and more recently in \cite{Jan,JHEP,Cost}.
%%%%%%%%%%%%%%%%%%%%%%%%%%%%%%%%%%%%%%%%%%%%%%%%%%%%%%%%%%%%%%%%%
\section *{Acknowledgments}
The authors would like to thank University of Tehran for supporting this project under the grants provided by the research council. M. N-Z also thanks H. Ramezani-Aval for useful discussions. 
%%%%%%%%%%%%%%%%%%%%%%%%%%%%%%%%%%%%%%%%%
\appendix
\section{Calculation of the inhomogeneous quasi-Maxwell equations in terms of the GEM field tensor}
In section III we gave an expression for the quasi-Maxwell form of the Einstein field equations in terms of the GEM field tensor in the coordinate system adapted to the TKVF. Here we find the 3-dimensional inhomogeneous quasi-Maxwell equations introduced in section II, starting with equation \eqref{inhomo},
\begin{equation}
\nabla_a {F_g}^{ab}=0.
\end{equation}
First we consider the spatial components of the field tensor,
\begin{align*}
\nabla_a {F_g}^{a\beta}&=0 \\
&=  \frac{1}{\sqrt{h} \sqrt{\gamma}}\dfrac{\partial}{\partial x^{a}} (\sqrt{h} \sqrt{\gamma} {F_g}^{a \beta})  \\
&=  \frac{1}{\sqrt{h} \sqrt{\gamma}}\dfrac{\partial}{\partial x^{\alpha}} (\sqrt{h} \sqrt{\gamma} (-h f^{\alpha \beta}))
\end{align*}
using $ f_{\alpha \beta}= \sqrt{\gamma} e_{\alpha \beta \gamma} {B_g}^{\gamma}$ yields 
\begin{align*}
\nabla_a {F_g}^{a\beta}&=0 \\
&= - \frac{1}{\sqrt{h} \sqrt{\gamma}}\dfrac{\partial}{\partial x^{\alpha}} (h^{3/2} e^{\alpha \beta \gamma} {B_g}_{\gamma})  \\
&= \frac{3h}{\sqrt{\gamma}} {E_g}_{\alpha} {B_g}_{\gamma} e^{\alpha \beta \gamma}- \frac{h}{\sqrt{\gamma}} e^{\alpha \beta \gamma} (\dfrac{\partial}{\partial x^{\alpha}} {B_g}_{\gamma}) \\
&= -3h (\textbf{{E}}_g \times \textbf{{B}}_g)^{\beta}+h (\nabla \times \textbf{{B}}_g)^{\beta}
\end{align*}
from which we obtain the following quasi-Maxwell equation in the three-space $ \Sigma_3 $
\begin{equation}
\nabla \times \textbf{{B}}_g= 3 \textbf{{E}}_g \times \textbf{{B}}_g \label{01}.
\end{equation}
Now we turn our attention to the spatio-temporal component of the equation, i.e, 
\begin{equation}
\nabla_a {F_g}^{a0}=0,
\end{equation}
which could be written as follows,
\begin{equation*}
\nabla_{\alpha} {F_g}^{\alpha 0}= \frac{1}{\sqrt{h} \sqrt{\gamma}}\dfrac{\partial}{\partial x^{\alpha}} (\sqrt{h} \sqrt{\gamma} {F_g}^{\alpha 0}).
\end{equation*}
Writing the GEM field tensor in terms of the GEM vector fields we have, 
\begin{equation}
\nabla_{\alpha} {F_g}^{\alpha 0}= \frac{2}{\sqrt{\gamma}}\dfrac{\partial}{\partial x^{\alpha}} (\sqrt{\gamma} {E_g}^{\alpha})- \frac{h}{\sqrt{\gamma}}\dfrac{\partial}{\partial x^{\alpha}} [\sqrt{\gamma} (\textbf{A}_g \times \textbf{B}_g)^{\alpha}]-[2{E_g}^{\alpha}-h(\textbf{A}_g \times \textbf{{B}}_g)^{\alpha}]{E_g}_{\alpha} \label{02}.
\end{equation}
in which for the second term in the right hand side we have
\begin{equation*}
\frac{h}{\sqrt{\gamma}}\dfrac{\partial}{\partial x^{\alpha}} [\sqrt{\gamma} (\textbf{A}_g \times \textbf{{B}}_g)^{\alpha}]= h {B_g}^2-h \textbf{A}_g \cdot(\nabla \times \textbf{{B}}_g)
\end{equation*}
so that we have 
\begin{equation*}
\nabla_{\alpha} {F_g}^{\alpha 0}=2 \nabla \cdot {E_g}-2{E_g}^2-h{B_g}^2+h \textbf{A}_g \cdot(\nabla \times \textbf{{B}}_g)+3h \textbf{{E}}_g \cdot(\textbf{A}_g \times \textbf{{B}}_g).
\end{equation*}
After making use of the equation \eqref{01} we end up with the desired result
\begin{equation*}
\nabla \cdot \textbf{{E}}_g={E_g}^2+\frac{1}{2}h{B_g}^2.
\end{equation*}
which is the other inhomogeneous quasi-Maxwell equation.
%%%%%%%%%%%%%%%%%%%%%%%%%%%%%%%%%%%%%%%%%%%%%%%%%%%%%%%%%%55
\section{Calculation of the extrinsic curvature in terms of the GEM fields}
Before proceeding with the calculation of the jump in the extrinsic curvature of the boundary hypersurface, it should be noted that  the following calculations are made in a gauge in which $A_{\alpha;\beta}+A_{\beta;\alpha}=0$. This is due to the fact that the combination $A_{\alpha;\beta}+A_{\beta;\alpha}$ is not an invariant under the gauge transformation $ {A_g}_\alpha \rightarrow {A_g}_\alpha + \partial_{\alpha}f $ representing the freedom in choosing the time origin \eqref{trans}, where it is known that all the 3-dimensional objects are scalars under the spacetime transformations \cite{Landau}. To keep with Landau's notation, in what follows we use $g^\alpha$ instead of ${A_g}^\alpha$ as the GEM 3-vector
potential. \\
We begin by writing all the connection coefficients in terms of the GEM 3-vector fields and then show that the two expressions \eqref{k1} and \eqref{k2} are equivalent in the coordinate system adapted to the TKVF and so being 3-tensorial relation on the hypersurface they are equal in all coordinate systems.\\
In terms of the metric components, the connection coefficients for a stationary spacetime (in the coordinate system adapted to the TKVF) are given by \cite{Landau},
\begin{gather}
\Gamma^{0}_{00} \doteq \frac{1}{2}g^\alpha h_{,\alpha}\nonumber\\
\Gamma^{0}_{\alpha0} \doteq \frac{h_{,\alpha}}{2h}+\frac{h}{2}g^{\beta}f_{\alpha\beta}-\frac{1}{2}g_\alpha g_\lambda h^{,\lambda}\nonumber\\
\Gamma^{0}_{\alpha\beta} \doteq -\frac{1}{2}(g_{\alpha,\beta}+g_{\beta,\alpha})-\frac{1}{2h}(g_{\alpha}h_{,\beta}+g_{\beta}h_{,\alpha})+g_{\delta}\lambda^{\delta}_{\alpha\beta}+\frac{1}{2}g_{\alpha}g_{\beta}g_{\delta}h^{,\delta}-\frac{h}{2}g^\lambda(g_{\alpha}f_{\beta\lambda}+g_{\beta}f_{\alpha\lambda})\nonumber\\
\Gamma^{\alpha}_{00} \doteq \frac{1}{2} h^{,\alpha}\nonumber\\
\Gamma^{\alpha}_{0\beta} \doteq \frac{h}{2}f_{\beta}^{~\alpha}-\frac{1}{2}g_{\beta}h^{,\alpha}\nonumber\\
\Gamma^{\alpha}_{\beta\gamma} \doteq \lambda^{\alpha}_{\beta\gamma}-\frac{h}{2}(g_{\beta}f_{\gamma}^{~\alpha}+g_{\gamma}f_{\beta}^{~\alpha})+\frac{1}{2}g_{\beta}g_{\gamma}h^{,\alpha}
\end{gather}
Rewriting the above equations in terms of the GEM 3-vector fields, they are given as follows,
\begin{gather}
\Gamma^{0}_{00} \doteq -h g^\alpha {E_g}_\alpha \nonumber\\
\Gamma^{0}_{\alpha0} \doteq -{E_g}_{\alpha}+\frac{h}{2}\sqrt{\gamma}g^{\beta}\epsilon_{\alpha\beta\gamma}{B_g}^{\gamma}+hg_{\alpha}g^{\lambda}{E_g}_{\lambda}\nonumber\\
\Gamma^{0}_{\alpha\beta}\doteq (g_{\alpha}{E_g}_{\beta}+g_{\beta}{E_g}_{\alpha})-hg_{\alpha}g_{\beta}g^{\lambda}{E_g}_{\lambda}-\frac{h}{2}g^{\gamma}(g_{\alpha}\sqrt{\gamma}\epsilon_{\beta\gamma\lambda}{B_g}^{\lambda}+g_{\beta}\sqrt{\gamma}\epsilon_{\alpha\gamma\lambda}{B_g}^{\lambda})\nonumber\\
\Gamma^{\alpha}_{00} \doteq -h\gamma^{\alpha\beta}{E_g}_\beta\nonumber\\
\Gamma^{\alpha}_{0\beta} \doteq \frac{h}{2}\sqrt{\gamma}{\epsilon_{\beta}^{~~\alpha}}_{\gamma}{B_g}^{\gamma}+hg_{\beta}\gamma^{\alpha\lambda}{E_g}_{\lambda}\nonumber\\
\Gamma^{\alpha}_{\beta\gamma} \doteq \lambda^{\alpha}_{\beta\gamma}-\frac{h}{2}(g_{\beta}\sqrt{\gamma}{\epsilon_{\gamma}^{~~\alpha}}_{\lambda}{B_g}^{\lambda}+g_{\gamma}\sqrt{\gamma}{\epsilon_{\beta}^{~~\alpha}}_{\lambda}{B_g}^{\lambda})-hg_{\beta}g_{\gamma}\gamma^{\alpha\lambda}{E_g}_{\lambda}
\end{gather}
Now what we need, is to show that the following relations hold between the connection coefficients and the components of the tensor $P^a_{bc}$,
\begin{eqnarray}
-\Gamma^{0}_{00}=P^0_{00} \;\; ; \;\;  -\Gamma^{0}_{\alpha0}=P^0_{\alpha0}\nonumber\\
-\Gamma^{0}_{\alpha\beta}=P^{0}_{\alpha\beta} \;\; ; \;\; -\Gamma^{\alpha}_{00}=P^{\alpha}_{00}\nonumber\\
-\Gamma^{\alpha}_{0\beta}=P^{\alpha}_{0\beta}  \;\; ; \;\; -\Gamma^{\alpha}_{\beta\gamma}+\lambda^{\alpha}_{\beta\gamma}=P^{\alpha}_{\beta\gamma}.
\end{eqnarray}
This could be achieved by calculating the components of $P^a_{bc}$ in the coordinate system adapted to the timelike Killing vector as follows,

\begin{gather}
P^0_{00} \doteq \frac{1}{h}{E_g}^0\xi_0\xi_0+\frac{1}{h}(-{E_g}_0^{~0}\xi_0)=hg^\alpha {E_g}_\alpha\nonumber\\
P^0_{\alpha0}\doteq \frac{1}{h}{E_g}^0\xi_\alpha\xi_0+\frac{1}{2h}(-{F_g}_\alpha^{~0}\xi_0-{F_g}_0^{~0}\xi_\alpha)={E_g}_{\alpha}-hg_{\alpha}g^{\beta}{E_g}_\beta-\frac{1}{2}h\sqrt{\gamma}g^{\beta}\epsilon_{\alpha \beta \lambda}{B_g}^{\lambda}\nonumber\\
P^{\alpha}_{00} \doteq \frac{1}{h}{E_g}^{\alpha}\xi_0\xi_0+\frac{1}{h}(-{F_g}_0^{~\alpha}\xi_0)=h\gamma^{\alpha\beta}{E_g}_\beta \nonumber \\
P^{\alpha}_{\beta0} \doteq \frac{1}{h}{E_g}^{\alpha}\xi_\beta\xi_0+\frac{1}{2h}(-{F_g}_\beta^{~\alpha}\xi_0-{F_g}_0^{~\alpha}\xi_\beta)=-g_{\beta}\gamma^{\alpha\lambda}{E_g}_\lambda-\frac{1}{2}h\sqrt{\gamma}{\epsilon_{\beta}^{~~\alpha}}_{\gamma}{B_g}^{\gamma}\nonumber\\
P^{0}_{\alpha\beta} \doteq \frac{1}{h}{E_g}^0\xi_\alpha\xi_\beta+\frac{1}{2h}(-{F_g}_\alpha^{~0}\xi_\beta-{F_g}_\beta^{~0}\xi_\alpha)  \nonumber \\ = -(g_{\alpha}{E_g}_{\beta}+g_{\beta}{E_g}_{\alpha})+hg_{\alpha}g_{\beta}g^{\lambda}{E_g}_{\lambda}+\frac{h}{2}g^{\gamma}(g_{\alpha}\sqrt{\gamma}\epsilon_{\beta\gamma\lambda}{B_g}^{\lambda}+g_{\beta}\sqrt{\gamma}\epsilon_{\alpha\gamma\lambda}{B_g}^{\lambda})\nonumber\\
P^{\alpha}_{\beta\gamma} \doteq \frac{1}{h}{E_g}^{\alpha}\xi_{\beta}\xi_\gamma+\frac{1}{2h}(-{F_g}_\beta^{~\alpha}\xi_\gamma-{F_g}_\gamma^{~\alpha}\xi_\beta) \nonumber \\
=\frac{h}{2}(g_{\beta}\sqrt{\gamma}{\epsilon_{\gamma}^{~~\alpha}}_{\lambda}{B_g}^{\lambda}+g_{\gamma}\sqrt{\gamma}{\epsilon_{\beta}^{~~\alpha}}_{\lambda}
{B_g}^{\lambda})+hg_{\beta}g_{\gamma}\gamma^{\alpha\lambda}{E_g}_{\lambda}
\end{gather}
comparing the above relations with those in $({\rm B}2)$ shows that the relations given in $({\rm B}3)$ are satisfied in the coordinate system adapted to the TKVF.
%----------------------------------------------------------------------------------------

%----------------------------------------------------------------------------------------
%	BIBLIOGRAPHY%----------------------------------------------------------------------------------------
%\pagebreak

%----------------------------------------------------------------------------------------

\end{document}